\documentclass{emulateapj}

\usepackage{verbatim}
\usepackage{amsmath}
\usepackage{hyperref}
\usepackage{breakurl}
\usepackage{float}

\usepackage{etoolbox}
\makeatletter
\patchcmd{\@verbatim}
  {\verbatim@font}
  {\verbatim@font\small}
  {}{}
\makeatother

\newcommand{\code}[1]{\texttt{#1}}

\begin{document}

\title{The Galaxy Cluster Merger Catalog: An Online Repository of Mock Observations from Simulated Galaxy Cluster Mergers}

\author{J. A. ZuHone\altaffilmark{1}, K. Kowalik\altaffilmark{2}, E. \"{O}hman\altaffilmark{3}, E. Lau\altaffilmark{3,4}, D. Nagai\altaffilmark{3,4,5}}

\altaffiltext{1}{Harvard-Smithsonian Center for Astrophysics, 60 Garden St., Cambridge, MA 02138, USA}
\altaffiltext{2}{National Center for Supercomputing Applications, University of Illinois at Urbana-Champaign, 1205 W. Clark St., MC-257, Urbana, IL 61801, USA}
\altaffiltext{3}{Department of Physics, Yale University, New Haven, CT 06520, USA}
\altaffiltext{4}{Yale Center for Astronomy and Astrophysics, Yale University, New Haven, CT 06520, USA}
\altaffiltext{5}{Department of Astronomy, Yale University, New Haven, CT 06520, USA}

\begin{abstract}
We present the ``Galaxy Cluster Merger Catalog.'' This catalog provides an extensive suite of mock observations and related data for N-body and hydrodynamical simulations of galaxy cluster mergers and clusters from cosmological simulations. These mock observations consist of projections of a number of important observable quantities in several different wavebands as well as along different lines of sight through each simulation domain. The web interface to the catalog consists of easily browseable images over epoch and projection direction, as well as download links for the raw data and a JS9 interface for interactive data exploration. The data is presented within a consistent format so that comparison between simulations is straightforward. All of the data products are provided in the standard FITS file format. Data is being stored on the yt Hub~(\url{http://hub.yt}), which allows for remote access and analysis using a Jupyter notebook server. Future versions of the catalog will include simulations from a number of research groups and a variety of research topics related to the study of interactions of galaxy clusters with each other and with their member galaxies. The catalog is located at \url{http://gcmc.hub.yt}.
\end{abstract}


\section{Introduction}\label{sec:intro}

Galaxy clusters are the largest gravitationally bound structures in the current universe. Originally identified as concentrations of galaxies in the optical, observations in the X-ray and millimeter wavelengths have revealed the bulk of baryonic material of clusters is comprised of a hot, magnetized plasma known as the intracluster medium \citep[ICM,][]{for72,sun72}. The kinetic energy of the galaxies and the temperature of the hot gas indicate that in order for the cluster to be gravitationally bound the majority of the mass must be in the form of cold dark matter \citep[CDM, first noted by][]{zwi37}.

Mergers between galaxy clusters represent the latest stage of cosmological structure formation. The most energetic events in the universe, mergers drive shocks and turbulence into the ICM, heating and stirring the cluster gas. These mergers also accelerate relativistic particles, which then produce radio relics and halos \citep{fer05,bru07,vwe10,bru12}. Cluster mergers have also revealed the different dynamical properties of the CDM, galaxies, and ICM, seen most vividly in the case of the Bullet Cluster \citep[][]{clo04,mar04}. Understanding cluster mergers is therefore vital to answering questions about the detailed physics of galaxy clusters as well as providing insights into the formation of cosmic structure.

The astrophysical literature is replete with simulations of galaxy cluster mergers, from binary merger simulations \citep[e.g.][]{ric01,poo06,zuh11a,don13} to studies of mergers in cosmological simulations \citep[e.g.][]{vaz09,ski13,yu15}. These simulations have often attempted to make predictions for what may be observed in a number of wavebands and made direct comparisons to observed merging systems. However, comparing the results of simulations of cluster mergers to these systems is often not straighforward; at what stage are we viewing the merger, and along what line of sight? To make matters more complicated, different combinations of merger epoch and line of sight can produce qualitatively similar projections of cluster emission, making it more difficult to use observations to determine these parameters and distinguish between different theoretical models.

We seek to address these issues by releasing the ``Galaxy Cluster Merger Catalog,'' an extensive suite of mock observations and related data for simulations of galaxy cluster mergers. We have produced 2D projections and slices of a number of different quantities relevant to observations of galaxy clusters from simulations of binary cluster mergers and cosmological simulations. These data products include clusters in various stages of merging and interaction with other systems viewed along a number of relevant projection axes. The data from the various simulations are presented within a consistent format so that comparison between simulations with different physics or different initial conditions is straightforward. The inclusion of data from both simulation types emphasizes the strengths of each: idealized merger simulations, due to their typically higher resolution and controlled set up, can provide insights into what kind of physics and merger configurations lead to certain observable features in cluster mergers. On the other hand, cosmological simulations capture more realistic mergers between clusters with substructure as well as cosmological accretion, providing a more direct comparison with observations.

The goal of this catalog is to provide a way to connect a variety of simulations of galaxy cluster mergers within a common interface with multiwavelength observations of real merging clusters, providing an opportunity for observers to compare clusters in their observations with a particular merging scenario. For example, the catalog may be used to interpret existing observations across multiple wavebands such as the Planck-Chandra Cluster Sample\footnote{\url{https://hea-www.harvard.edu/CHANDRA_PLANCK_CLUSTERS/}} \citep{and17}, or be analyzed to make predictions for what various investigations of real clusters may reveal for future observations and missions, such as {\it Athena}\footnote{\url{http://www.the-athena-x-ray-observatory.eu/}} and {\it Lynx}\footnote{\url{https://wwwastro.msfc.nasa.gov/lynx/}}.

\section{The Data}

\subsection{The Simulations}

At the time of writing, the catalog contains data products from idealized binary cluster merger simulations and one cosmological simulation. The original works from which these simulations are drawn, in addition to some details of the simulations, are listed in Table \ref{tab:sim_table}. The types of data products listed here are appropriate for such simulations. In the future, the catalog will include data products from additional simulations of idealized binary mergers, cosmological simulations, and simulations of interactions between clusters and their member galaxies. The types of data products are likely to expand and diversify with the inclusion of new simulations. For more discussion of future directions for the catalog and long-term issues, see Section \ref{sec:future}.

\begin{table*}
\caption{Original Works Pertaining to Catalog Data\label{tab:sim_table}}
\begin{center}
\begin{tabular}{llll}
\hline
\hline
Original Reference & \# of Simulations & Code & Basic Physics \\
\hline
\citet{zuh10} & 4 & FLASH & N-body/Hydrodynamics \\
\citet{zuh11a} & 9 & FLASH & N-body/Hydrodynamics \\
\citet{zuh11b} & 6 & FLASH & Magnetohydrodynamics \\
\citet{zuh15} & 9 & Athena & Magnetohydrodynamics \\
\citet{nel14} & 1 & ART & N-body/Hydrodynamics \\
\hline
\end{tabular}
\end{center}
\end{table*}

The galaxy cluster merger simulation data in the catalog comes from state-of-the-art N-body and hydrodynamics codes such as FLASH \citep{dub09}, ART \citep{kra99, kra02, rud08}, and Athena \citep{sto08}. The exact physics and algorithms employed by the simulations vary, but in general:

\begin{itemize}
\item Each simulation is simulated using an adaptive or static mesh refinement (AMR/SMR) grid \citep{ber89} with varying resolution throughout the domain, with refinement occuring on criteria such as sharp jumps in density and temperature, matter density, and selected regions such as the cluster center.
\item The equations of hydrodynamics (HD) or magnetohydrodynamics (MHD) are modeled using a conservative finite-volume scheme, employing Riemann solvers for evolving the flux of physical quantities and using high-order reconstruction schemes such as the Piecewise-Parabolic Method \citep[PPM,][]{col84}. If present, magnetic fields are evolved such that the condition $\nabla \cdot {\bf B} = 0$ is met, typically by a constrained transport scheme \citep[CT,][]{eva88}.
\item Each simulation assumes an ideal gas law equation of state with $\gamma = 5/3$ and primordial abundances of H/He with trace amounts of metals, yielding a mean molecular weight of $\mu \simeq 0.6$.
\item If dark matter is included, it is modeled by an N-body solver for a collection of collisionless massive particles, which interact with the gas component only via gravity.
\item The gravity in the simulations is either modeled as a rigid gravitational potential associated with each cluster or by computing the self-gravity of the gas and dark matter using a Poisson solver \citep[e.g.,][]{ric08}.
\item Cosmological simulations in the catalog include the effect of cosmological expansion by evolving the equations of hydrodynamics and particle advancement in comoving coordinates and with terms including the dependence on the scale factor of the universe.
\item Depending on the goals of the simulation study, other physics, such as viscosity, thermal conduction, radiative cooling, star formation and energy feedback from black holes, etc., may be included.
\end{itemize}

Though currently the simulations represented in the catalog are of the grid-based variety, eventually the catalog will also include data from smoothed particle hydrodynamics (SPH) and moving-mesh simulations.

\subsection{Methods and Technologies}\label{sec:methods}

In this section, we briefly outline the methods and software used to generate the data products from the original simulation data and host them on the catalog website.

\subsubsection{Python Scientific Software}\label{sec:software}

The projected data is generated from the original simulation data using \code{yt} \citep{tur11}, an open-source Python package for analyzing and visualizing volumetric data. \code{yt} has capabilities for slicing and projecting physical fields from multi-resolution datasets such as those from galaxy cluster merger simulations along arbitrary lines of sight. Through analysis modules and affiliated packages, \code{yt} also has capabilities to produce synthetic Sunyaev-Zeldovich (S-Z) and X-ray observations. We provide details on how this is done in Sections \ref{sec:sz} and \ref{sec:xray}.

\code{AstroPy} \citep{apy13} is used to take the projected data generated in \code{yt} and produce data outputs in the standard Flexible Image Transport System (FITS) data format \citep{pen10} and generate coordinates for the files using the World Coordinate System \citep[WCS,][]{gre02, cal02} specification.

\subsubsection{JS9}\label{sec:js9}

JS9 (\url{http://js9.si.edu}) is a browser-based implementation of SAOImage DS9 (\url{http://ds9.si.edu/}), an astronomical imaging and data visualization application. JS9 provides interactive exploration of FITS image and table data including zooming, panning, colormaps, scaling, and region files. On each ``epoch page'' (Section \ref{sec:epoch_page}) there is an interface to JS9 and links to open FITS files from that epoch in the JS9 interface.

\subsubsection{The yt Hub}\label{sec:yt_hub}

The catalog data is stored on the \code{yt Hub}\footnote{\url{https://hub.yt/}}, a conglomerate of various open source
services hosted at the National Center for Supercomputing Applications (NCSA), creating an environment that allows one to collect, remotely analyze, share and publish scientific datasets. At its core, the \code{yt Hub} uses \code{Girder}\footnote{\url{https://girder.readthedocs.io}}, a data management platform that allows to transparently store, serve, and proxy data from heterogeneous backend storage engines through a single
web API. Data served by \code{Girder} can easily be gathered into dynamical hierarchies such as collections, folders, and
items. This data management system is integrated with \code{JupyterHub}\footnote{\url{https://jupyterhub.readthedocs.io/}} and allows authenticated users to spawn Jupyter notebook servers with direct access to the data that is provided as a dynamically composed FUSE filesystem. Any Jupyter
notebook created in that environment during an interactive session is safely stored back in the \code{Girder} instance,
preserving the user's analysis workflow and increasing research reproducibility. In the future, we plan to have this
notebook capability integrated directly with the catalog website.

\subsection{The Data Products}\label{sec:data}

The data products currently provided in the catalog are 2D selections or reductions of the original 3D datasets, such as slices or projections. The particular products included for a particular simulation depend on its physical and algorithmic details (e.g., slices of the magnetic field strength and Faraday rotation measure projections are only included for MHD simulations).

The data products are all written in the FITS format, in image and table form. Each FITS file may contain more than one image or table extension. The header of each extension contains coordinate information stored using standard WCS keywords. Each FITS file contains one or both of the following two coordinate systems:

\begin{itemize}
\item A linear coordinate system which corresponds to the coordinate system of the original dataset. The length units are in kpc. For most of the FITS files, this is the first and primary WCS (e.g., the one that appears by default in ds9).
\item A celestial coordinate system in RA and Dec using the tangential projection. The angle units are in degrees. For most of the FITS files, this is the secondary WCS (e.g., ``WCS a'' in ds9).
\end{itemize}

For example, a header for one of the FITS images corresponding to a projected quantity may look like this (omitting some keywords for clarity):

\begin{verbatim}
NAXIS   =                    2
NAXIS1  =                 2048
NAXIS2  =                 2048
EXTNAME = 'KT      '
BTYPE   = 'kT      '
BUNIT   = 'keV     '
WCSAXES =                    2
CRPIX1  =               1024.5
CRPIX2  =               1024.5
CDELT1  =     0.97653794699453
CDELT2  =     0.97653794699453
CUNIT1  = 'kpc     '
CUNIT2  = 'kpc     '
CTYPE1  = 'LINEAR  '
CTYPE2  = 'LINEAR  '
CRVAL1  =                  0.0
CRVAL2  =                  0.0
LATPOLE =                 90.0
WCSNAME = 'yt      '
WCSAXESA=                    2
CRPIX1A =               1024.5
CRPIX2A =               1024.5
CDELT1A = -0.00028118222874698
CDELT2A =  0.00028118222874698
CUNIT1A = 'deg     '
CUNIT2A = 'deg     '
CTYPE1A = 'RA---TAN'
CTYPE2A = 'DEC--TAN'
CRVAL1A =                 30.0
CRVAL2A =                 45.0
LONPOLEA=                180.0
LATPOLEA=                 45.0
WCSNAMEA= 'celestial'
RADESYSA= 'ICRS    '
TIME    =    1.300254073176463
\end{verbatim}

It can be seen here that the default WCS, \code{WCSNAME = 'yt'}, is in linear coordinates, and the second WCS, \code{WCSNAMEA = 'celestial'}, is in celestial coordinates. The relationship between the two depends on the angular diameter distance to the source, which depends on the redshift and the given cosmology. The pixel scale of each FITS file is equivalent to the finest cell size of the simulation.

With the exception of the X-ray event files (Section \ref{sec:xray}), no attempt is made in the mock observations to include statistical and systematic errors on the observed quantities, instrumental effects, or background contamination. It is not possible to anticipate in advance what assumptions the end-user may want to make about these issues, so these effects have been left out so they may be added on a case-by-case basis for different situations.

The next several subsections describe in detail the types of data products.

\subsubsection{Slices}\label{sec:slices}

Each simulation epoch has a FITS file containing slices through the central merger plane of a subset of the following fields, contained within different HDUs with the following extension names:

\begin{itemize}
\item \code{"density"}: Gas density in units of $\rm{M_\odot~kpc^{-3}}$.
\item \code{"dark\_matter\_density"}: Dark matter density in units of $\rm{M_\odot~kpc^{-3}}$.
\item \code{"kT"}: Gas temperature in units of keV.
\item \code{"velocity\_x"}: The x-component of the gas velocity in units of $\rm{km~s^{-1}}$.
\item \code{"velocity\_y"}: The y-component of the gas velocity in units of $\rm{km~s^{-1}}$.
\item \code{"magnetic\_field\_strength"}: The magnetic field strength in units of $\mu$G.
\end{itemize}

For a number of the simulations, fields representing the mass fraction of gas from each cluster within a particular cell are also included. The header for each image includes a linear WCS with distances given in kpc. These slices can be used to examine the history of the merger from the perspective of the 3D physical quantities which are evolved by the simulation, and from which the 2D projected quantities are derived.

\subsubsection{Projections}\label{sec:projections}

Projections are taken along several lines of sight. Currently, these include the three major axes of the simulation domain: $x$, $y$, and $z$. In the future, projections along off-axis directions will be added. For distance and redshift-dependent quantities, each simulation employs either the cosmology and redshift from the simulation, or for simulations without such information, a redshift and a $\Lambda{\rm CDM}$ cosmology is assumed with the following parameters: $H_0 = 71~{\rm km~s^{-1}~Mpc^{-1}},$ $\Omega_m = 0.27$, and $\Omega_\Lambda = 0.73$. The projection files each have a subset of the following fields contained within different HDUs with the following extension names:

\begin{itemize}
\item \code{"xray\_emissivity"}: X-ray photon surface brightness in the 0.5-7.0~keV (observer) band, computed using an APEC model, in units of ${\rm photons~s^{-1}~{cm}^{-2}~{arcsec}^{-2}}$. For simulations without metallicity, a spatially constant metallicity of ${\rm Z = 0.3~Z_\odot}$ is assumed.
\item \code{"kT"}: Emission-weighted projected temperature, using the emissivity described above, in units of keV.
\item \code{"total\_density"}: Total mass density (gas and dark matter) in units of $\rm{M_\odot~{kpc}^{-3}}$.
\item \code{"szy"}: The integrated ``y-parameter'' for the thermal S-Z effect, given by
\begin{equation}
y_{\rm tSZ} = \int{\frac{k_BT}{m_e{c^2}}\sigma_T{n_e}{\rm d\ell}}.
\end{equation}
\item \code{"sz\_kinetic"}: The integrated ``y-parameter'' for the kinetic S-Z effect, given by
\begin{equation}
y_{\rm kSZ} = \int{\frac{v_\ell}{c}\sigma_T{n_e}{\rm d\ell}}.
\end{equation}
\item \code{"rotation\_measure"}: Faraday rotation measure in units of rad~m$^{-2}$, given by:
\begin{equation}
{\rm RM} = \frac{e^3}{2\pi{m_e}^2c^4}\int{n_e}{B_{\ell}}{\rm d}{\ell}.
\end{equation}
\end{itemize}

\subsubsection{Sunyaev-Zeldovich Effect Projections}\label{sec:sz}

Projections of the full S-Z signal are also computed for some simulations, using the \code{SZpack}\footnote{\url{http://www.cita.utoronto.ca/~jchluba/Science_Jens/SZpack/SZpack.html}} library \citep{chl12, chl13} to compute the S-Z signal, including thermal and kinetic contributions as well as relativistic corrections. More details on how these projections were computed can be found at \url{http://yt-project.org/doc/analyzing/analysis_modules/sunyaev_zeldovich.html}. They are stored in separate FITS files from the other projections. Currently, these files contain a subset of these fields with the follwing extension names:

\begin{itemize}
\item \code{"Tau"}: Compton optical depth of the cluster gas.
\item \code{"TeSZ"}: Mass-weighted projected temperature in units of keV.
\item \code{"90\_GHz"}: S-Z signal at 90~GHz in units of MJy~sr$^{-1}$.
\item \code{"180\_GHz"}: S-Z signal at 180~GHz in units of MJy~sr$^{-1}$.
\item \code{"240\_GHz"}: S-Z signal at 240~GHz in units of MJy~sr$^{-1}$.
\end{itemize}

The computed S-Z signal at other frequencies may be added in later revisions of the catalog.

\subsubsection{X-ray Event Files}\label{sec:xray}

The X-ray emissivity projection listed above represents a single broad energy band from 0.5-7.0 keV. However, to do analysis of the thermal and compositional properties of the ICM, high-resolution energy spectra are required. Instead of providing similar maps at the same spatial resolution for hundreds or even thousands of energy bins for every epoch and projection direction, which would be prohibitive in terms of disk space, the catalog provides event files.

These X-ray event files are standard products which can be manipulated and analyzed with common X-ray analysis tools, such as ds9, \code{CIAO}, and the \code{HEASOFT} software suite. The events themselves have been generated from the original 3-D datasets using the \code{pyXSIM}\footnote{\url{http://hea-www.cfa.harvard.edu/~jzuhone/pyxsim}} package \citep{zuh14}, which not only generates the 2-D projected X-ray events but can also convolve the photons with instrumental response matrices. At the time of writing, event files for the {\it Chandra} ACIS-I detector are provided, but eventually event files for other instruments will be included, as well as raw, unconvolved events which may be used as input for any instrument simulator.

The event files are designed to provide approximate representations of real observations, with some simplifications. First, the event energies have been convolved with the ACIS-I on-axis responses over the entire field of view. Second, the pixel size corresponds to the width of the finest simulation cell size (in line with the other data products), instead of the pixel scale of the detector. Thirdly, since the {\it Chandra} PSF is very small and for the distances assumed here the cell sizes are typically much larger, no PSF effects have been applied. The idea is not to create an exact simulation, but instead to approximate the statistical properties of the source while allowing one to use standard X-ray tools to analyze the spectral and spatial properties of the data.

\subsubsection{Mock Galaxy Catalogs}\label{sec:galaxies}

\begin{figure*}
\begin{center}
\includegraphics[width=0.9\textwidth]{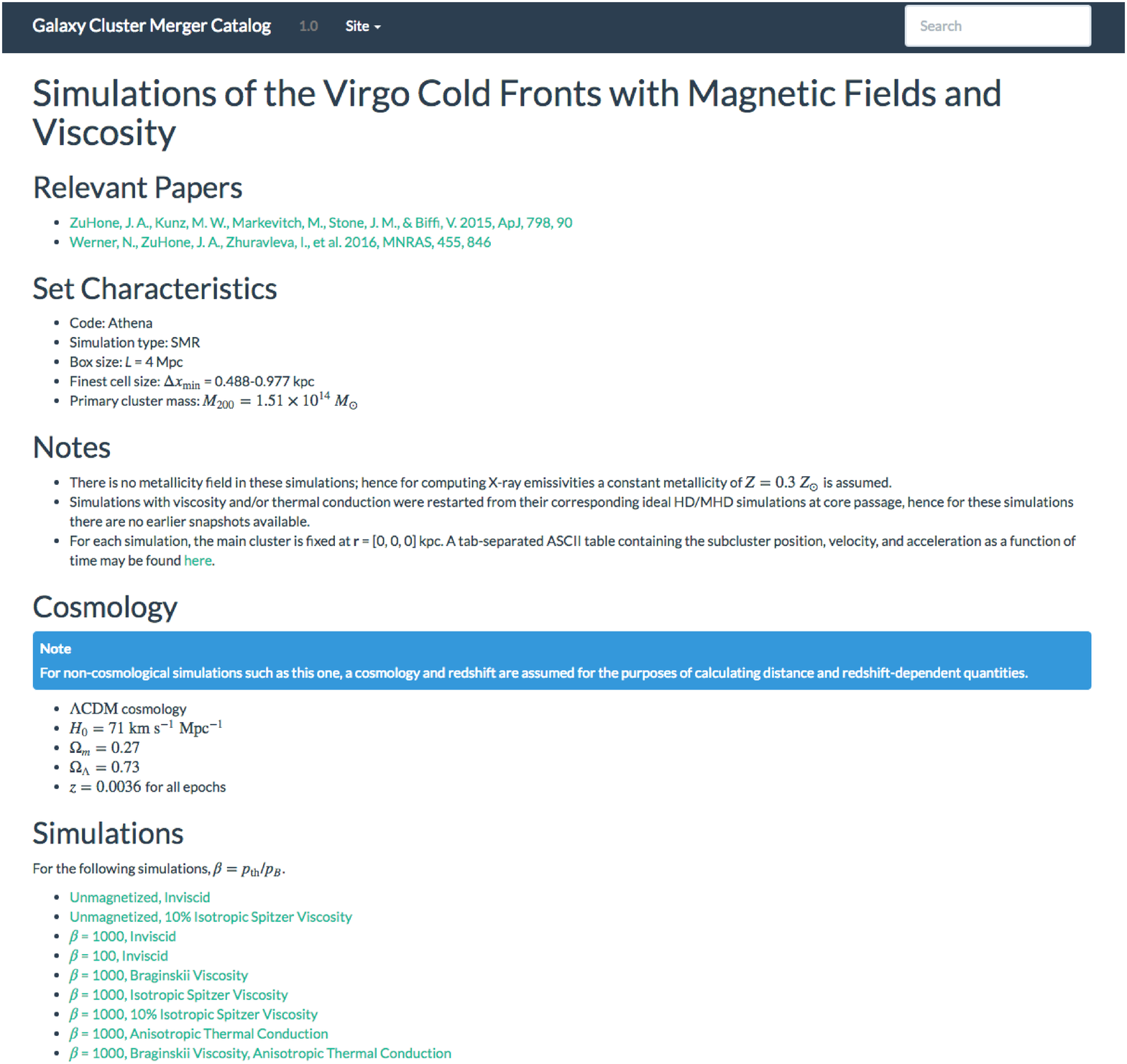}
\caption{An example ``set page.'' Links are provided to original journal articles, various simulation parameters and other information are tabulated, and links to the data for the individual simulations appear here. This particular page is linked at \url{http://gcmc.hub.yt/virgo/index.html}.\label{fig:set_page}}
\end{center}
\end{figure*}

\begin{figure*}
\begin{center}
\includegraphics[width=0.9\textwidth]{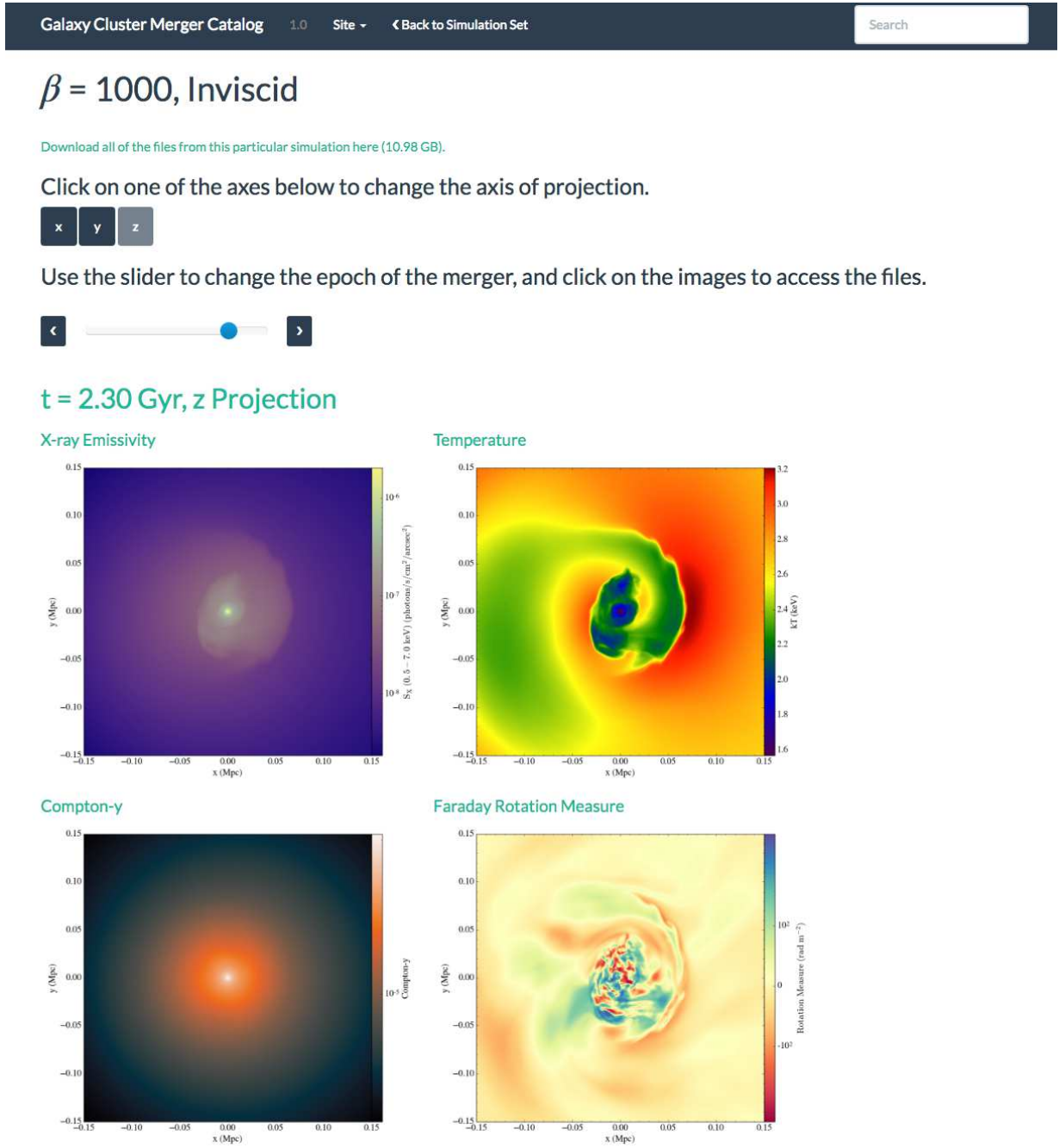}
\caption{An example ``simulation page.'' Slider bars allow exploration of different epochs, and buttons near the top choose different projection axes. The figures link to the page for the files for the corresponding epoch of the simulation. This particular page is linked at \url{http://gcmc.hub.yt/virgo/novisc/index.html}.\label{fig:sim_page}}
\end{center}
\end{figure*}

\begin{figure*}
  \begin{center}
  \includegraphics[width=0.9\textwidth]{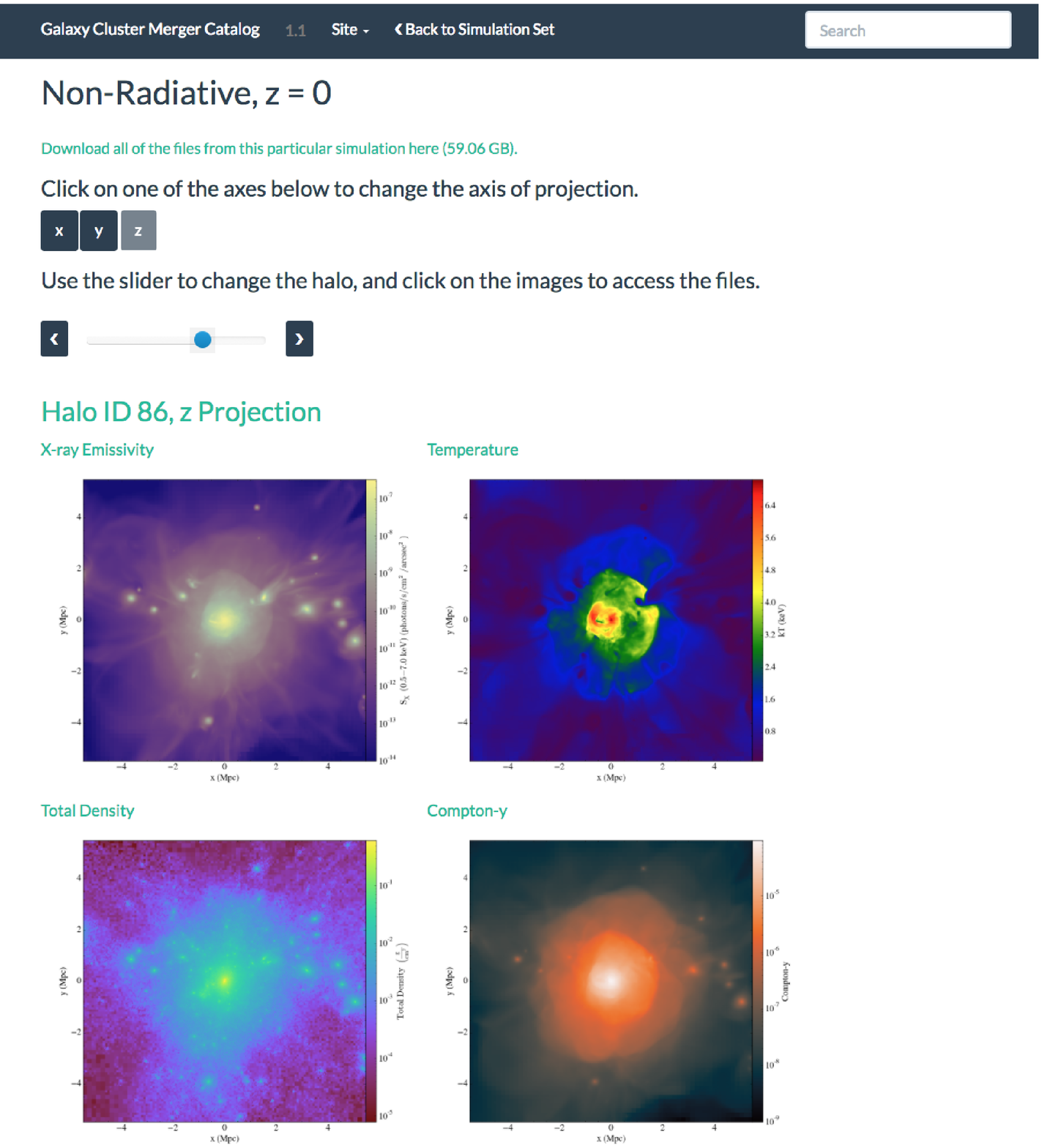}
  \caption{Another example of a ``simulation page,'' except showing halos from a cosmological simulation instead of epochs. Slider bars allow exploration of different halos, and buttons near the top choose different projection axes. The figures link to the page for the files for the corresponding halo ID. This particular page is linked at \url{http://gcmc.hub.yt/omega500/non_radiative_1.0005/index.html}.\label{fig:halos_page}}
  \end{center}
  \end{figure*}
  
\begin{figure*}
\begin{center}
\includegraphics[width=0.9\textwidth]{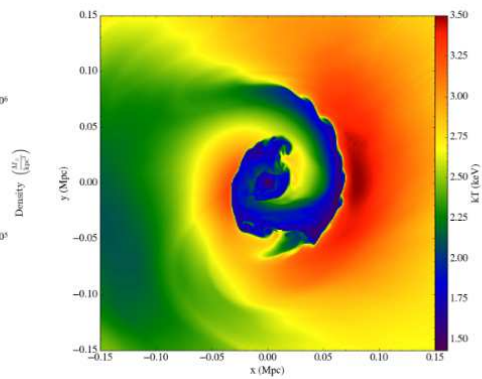}
\caption{An example ``epoch page.'' Links are provided to similar simulations with different physics at the same epoch at the top of the page. For each data type, figures of some of the fields are shown and links to download the data or display in JS9 are provided. This particular page is linked at \url{http://gcmc.hub.yt/virgo/novisc/0046.html}.\label{fig:epoch_page}}
\end{center}
\end{figure*}

\begin{figure*}
\begin{center}
\includegraphics[width=0.9\textwidth]{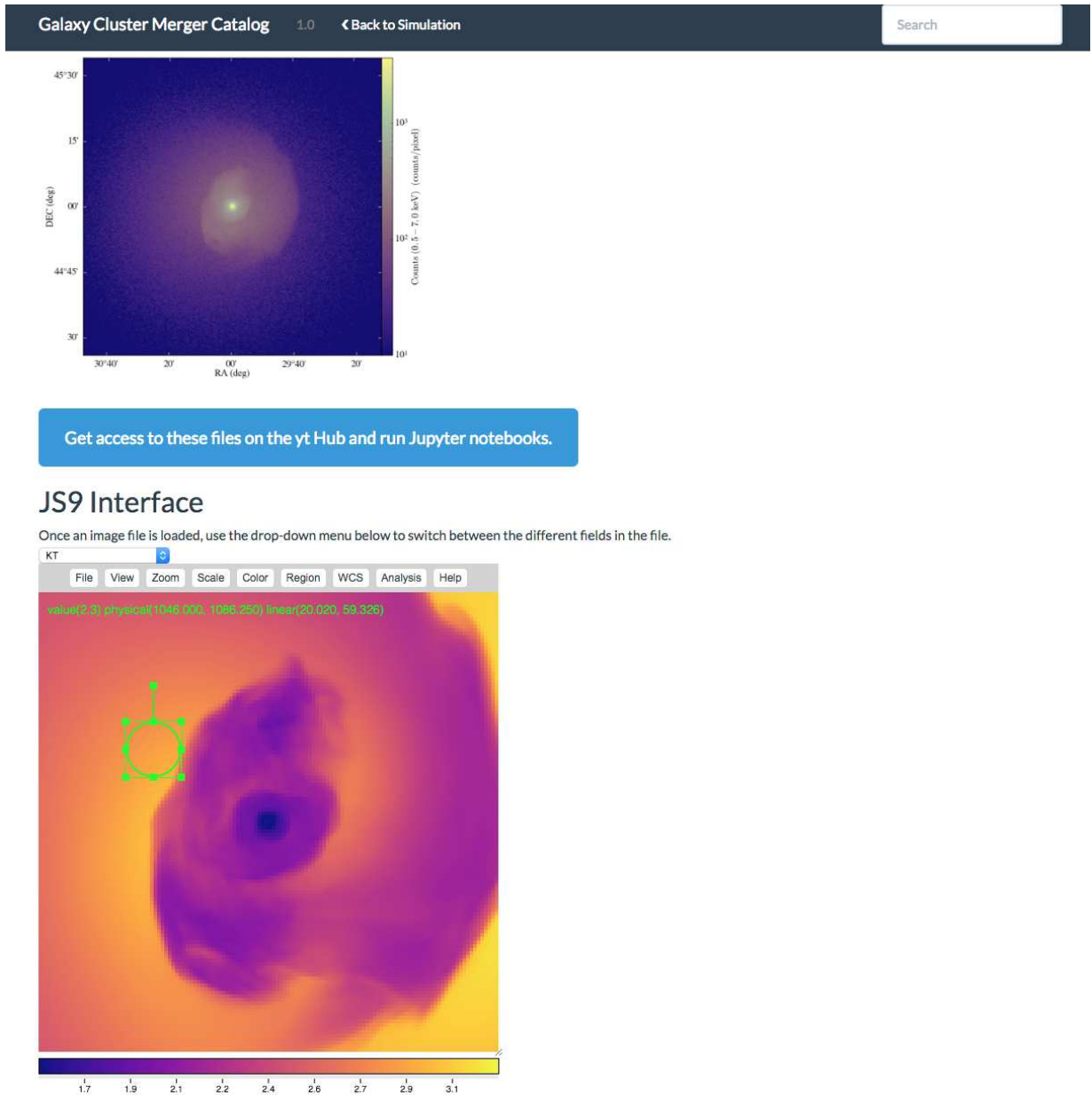}
\caption{The bottom of an ``epoch page,'' showing the simulated X-ray event file, the link to the Jupyter notebook directions, and the JS9 interface with a projection file loaded, and a circular region selected. This particular page is linked at \url{http://gcmc.hub.yt/virgo/novisc/0046.html}.\label{fig:epoch_page2}}
\end{center}
\end{figure*}

\begin{figure*}
\begin{center}
\includegraphics[width=0.95\textwidth]{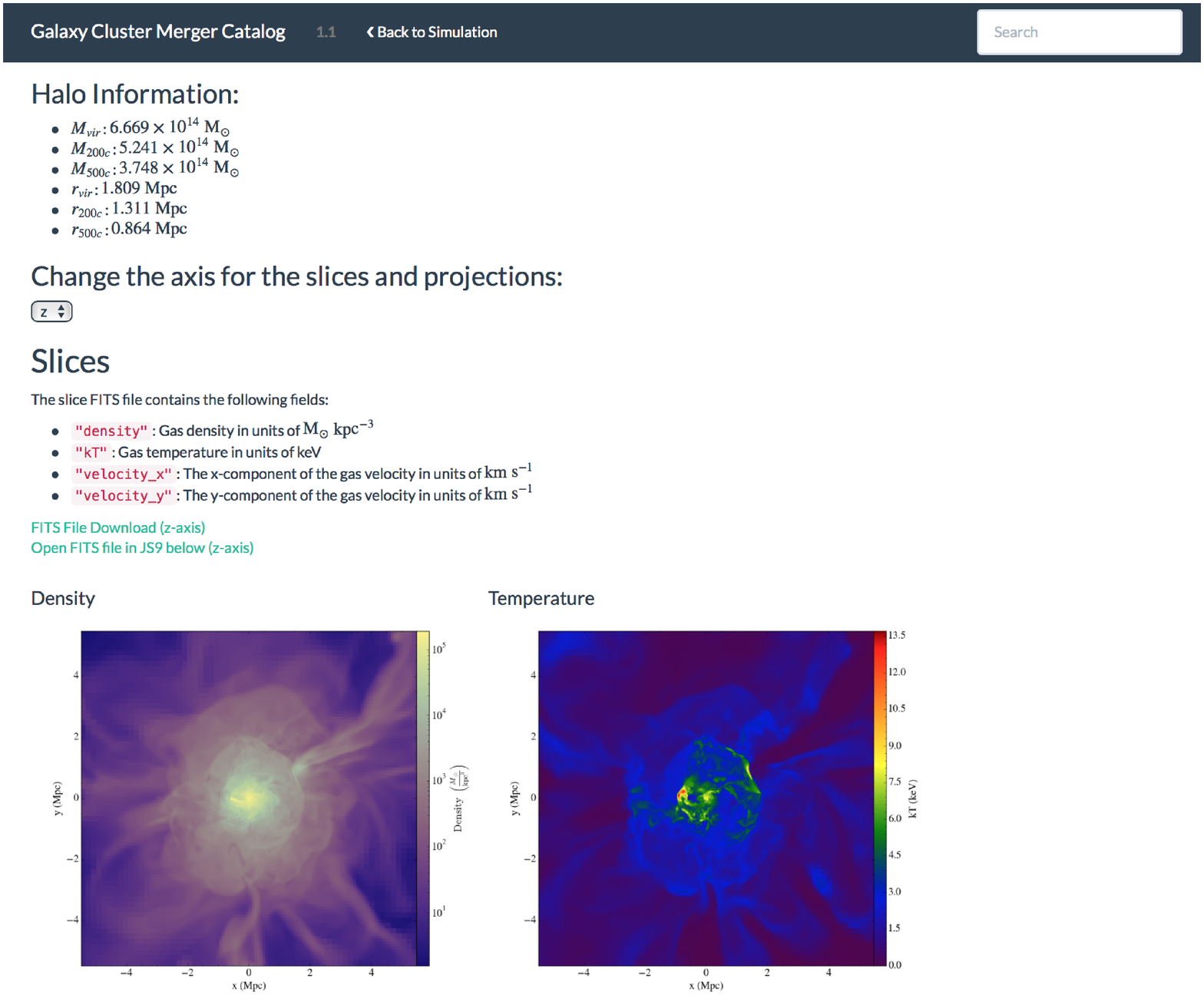}
\caption{A portion of a ``halo page,'' showing the details of the halo information and the slice data for this particular halo. This particular page is linked at \url{http://gcmc.hub.yt/omega500/non_radiative_1.0005/00086.html}.\label{fig:halo_page}}
\end{center}
\end{figure*}

A number of the simulations have been run using an N-body code to represent the collisionless dark matter component of the cluster via massive particles. A number of these dark matter particles can be sampled from the simulation to represent mock ``galaxies,'' which can also be treated as a collisionless component. For some of these simulations, we have created mock galaxy catalogs, which consist of the sky positions and redshifts of ``galaxy'' particles.

A number of particles are drawn from each halo within a radius of $r_{200}$, determined by the mass-richness relation of \citet{for14}, from the intitial simulation epoch, which are the ``galaxies.'' The same galaxy particles are tracked for the rest of the epochs in the simulation. The FITS binary tables for the galaxies include sky positions, line-of-sight velocities, an identifier for which halo each galaxy originally belonged to, and unique IDs for every galaxy. These particles provide a way of measuring the kinematics of the merger from the perspective of the collisionless material with a statistical significance that is comparable to that obtained from measured redshifts of member galaxies in real clusters. No redshift errors have been applied to the galaxy velocities, which is an exercise left to the end-user.

A ds9 region file containing point regions corresponding to the galaxy positions is provided for each epoch and line-of-sight in addition to the FITS file, which allows the galaxy positions to be plotted over the projections of the other fields. At the time of writing, mock galaxy catalogs have been created for the ``A Parameter Space Exploration of Galaxy Cluster Merger'' simulations found at \url{http://gcmc.hub.yt/fiducial/index.html}, an example of which may be found at \url{http://gcmc.hub.yt/fiducial/1to1_b0.5/0139.html}, under the heading ``Galaxies''.

\subsection{The Interface}\label{sec:interface}

The web interface to the catalog is designed to provide straightforward interactive browsing of a given simulation's data across its many epochs and between several projection axes. This facilitates comparisons between the mock observations and actual observations of clusters so that a potential analog of a particular system may be found easily. Like the data itself, the web interface is also hosted at NCSA.

The portion of the website containing the simulation data is structured using the a hierarchy of pages. Simulations are grouped together in ``sets'', which are simulations with shared characteristics. The page for each simulation within a set has links to pages for the different epochs (or individual cluster halos in the case of a cosmological simulation). Finally, the page for each epoch or halo within a simulation provides the access to the corresponding data products that were described above. We now describe these pages in detail.

\subsubsection{The Set Page}\label{sec:set_page}

The highest level of organization within the catalog is that of a ``set'' of simulations, which are simulations with shared characteristics. For example, these may be simulations with identical initial conditions but different input physics (e.g., pure hydrodynamics versus MHD, or a non-radiative cosmological simulation versus a simulation with star formation and feedback), or simulations with identical physics but different initial conditions (e.g., varying the mass ratio and/or impact parameter in binary merger simulations). Simulations within a set were run with the same code. The ``set page'' describes the general characteristics of a single set of simulations from a particular investigation and provides links to the data for the individual simulations. This page also provides links to journal articles which the simulation data appeared in, as well as basic characteristics of the simulation data including the size of the domain, the finest cell size, the characteristic mass of the clusters, the cosmological parameters, etc. This page may also include links to supplementary data files that are common to all of the simulations in the set.

All of the set pages are linked from the main simulation page, which can be found at \url{http://gcmc.hub.yt/simulations.html}. Figure \ref{fig:set_page} shows an example of a set page.

\subsubsection{The Simulation Page}\label{sec:sim_page}

Every simulation of a set has a ``simulation page'' which shows various quantities projected perpendicular to the merger plane for every epoch or halo of the simulation that is available. The different epochs or halos can be navigated using a slider bar, and buttons above the slider bar can be used to look at the same images from a different line of sight. The links attached to the images lead to the epoch or halo pages for this simulation. Simulation pages for a binary merger simulation with different epochs and a cosmological simulation with different halos are shown in Figures \ref{fig:sim_page} and \ref{fig:halos_page}.

\subsubsection{The Epoch and Halo Pages}\label{sec:epoch_page}

Any given link on the simulation page leads to an ``epoch page'' or ``halo page'' where images from a particular simulation epoch for idealized simulations, or a particular halo for cosmological simulations, can be examined and the corresponding data may be obtained. Links are provided to similar simulations with different input physics at the same epoch or for the same halo. This page also contains information about the different types of data within the files. A drop-down menu is provided for switching between the different projection axes available. Examples of epoch and halo pages are shown in Figures \ref{fig:epoch_page}-\ref{fig:halo_page}.

A link is provided on each epoch or halo page to the location of the files on the \code{yt Hub}, so that one may start Jupyter notebooks with access to the data. Each epoch page also contains a JS9 application for opening the various FITS files associated with that particular epoch or halo. A drop-down menu is provided for navigating between the various fields in each FITS file. 

\section{Future Directions}\label{sec:future}

There are a number of extensions to the catalog and its capabilities that we envision for the future, which we briefly discuss in this section.

\subsection{Adding New Simulations}\label{sec:new_sims}

The catalog is designed to be easily expanded to include new simulation data products related to galaxy cluster mergers or the more general arena of galaxy clusters, groups, or galaxies in the cosmological context. One way to encourage the inclusion of new simulations in the catalog is to provide Python scripts which may be used to generate the data products, including FITS image and table files and PNG images. A future version of the catalog will include a suite of such scripts which will be publicly available for this purpose, and directions for how to upload the resultant products to the \code{yt Hub}.

\subsection{Adding New Data Products}\label{sec:new_data}

The current data products include slices and projections of various physical quantities, but they have a distinct tilt towards X-ray observations of clusters due to the focus of the simulations included so far on the physics of the ICM. Many other simulations of clusters available in the literature have additional physics included in the simulation, such as star formation, active galactic nuclei, and relativistic particle acceleration. These physical processes produce observations in other wavebands, such as the optical, ultraviolet, or radio, which could be included as simulated observations in the form of new FITS image or table types in the catalog.

\subsection{Sustainability}\label{sec:sustainability}

Maintaining such a large catalog of data, whether from simulations or otherwise, is a substantial undertaking. It first and foremost requires that the computational resources necessary to store the data and serve it to users will be maintained and funded for the forseeable future. The website \url{http://gcmc.hub.yt} itself (including the widgets, documentation, and user interface) is completely separated from the data itself, which is hosted on the yt Hub. Therefore it is completely portable, and may be hosted virtually anywhere, provided proper configuration of Cross-Origin Resource Sharing (CORS) between it and the \code{yt Hub}.

As for the \code{yt Hub} itself, its infrastructure and software is an integral part of other NCSA projects related to data publication and data provenance\footnote{One such example being WholeTale, \url{https://www.nsf.gov/awardsearch/showAward?AWD_ID=1541450}}, which we hope makes it sustainable for the foreseeable future.

The second important resource is of course that of the human beings who are maintaining and updating the catalog. The authors themselves are committed to the long-term development and maintenance of the various components required to sustain this project, and our hope is that by getting more researches involved by including their data we can make the catalog a project which is sustained by a wider community. As a first step toward this goal, the code for the website is open-source and hosted at \url{http://github.com/jzuhone/cluster_merger_catalog}. Next steps include streamlining the code for the website itself and providing templates to simplify the generation of pages for new simulations, including directions for how to create a new set of pages for a simulation and how to submit them as a pull request to the repository.

\section{Summary}\label{sec:summary}

We have presented the ``Galaxy Cluster Merger Catalog,'' a website designed to provide a simple yet powerful interface to accessing mock observations of galaxy clusters. The catalog includes a suite of mock observations from a number of binary cluster merger and cosmological simulations, with different initial conditions and/or input physics. In particular, the catalog provides a number of X-ray and S-Z synthetic observations as well as mock galaxy catalogs for some simulations. Projections of the observables have been taken along a number of lines of sight through the simulation box, and at a large number of epochs for each simulation so that the merger history is well-sampled in time. The catalog interface is designed to facilitate quick and straightforward comparisons between the different simulations. This large parameter space of cluster mergers will be useful when comparing with observed cluster mergers in a number of wavebands.

The catalog is designed so that it may be easily extended and expanded. Future goals include:

\begin{itemize}
\item Adding cluster merger simulations from other research groups
\item Adding mock observations from other wavebands, such as radio and optical
\item Adding additional lines of sight through the simulation domain
\item Implementing more ways to interact with the data, such as Jupyter notebooks that may be run on the server
\item Adding other types of data products, including movies, unconvolved X-ray event lists, and the original 3D datasets
\item Providing example scripts to show other researchers how to generate similar data products from their simulations
\end{itemize}

At the time of writing, the catalog includes data products from binary merger simulations carried out by the first author, and from a cosmological simulation carried out by \citet{nel14}. The catalog is released and designed with the intent that it may be expanded to include simulation data products related to galaxy cluster mergers or the broader subject of cosmological structure formation from other works and authors. We have already received a positive response from the community in this regard and have plans underway to incorporate new simulations into the catalog in the near future.

Comparisons of data from the catalog to current observations have already been made to useful effect by \citet{wal17} and Douglass et al. 2017, in preparation. The catalog is also being explored as an avenue for further simulation investigations as well. The simulated X-ray images from several of the simulations are currently being analyzed with edge-detection techniques to determine if such an approach can reveal differences in the X-ray surface brightness distribution produced by varying the plasma physics of the ICM, including the strength of the magnetic field and viscosity (Bellomi et al., in preparation). Another area of research that is being explored by the authors of this work involves extracting cluster merger parameters from mergers in cosmological simulations and using them as inputs for idealized simulations, applying different prescriptions for the physics of the ICM in different simulations. This will be particularly useful from the perspective of determining the effect of varying the ICM physics on the velocity field of merging clusters, which will be an area of intense research with the advent of X-ray observatories carrying instruments which can measure such motions. The fruits of this research will eventually become part of the catalog as well. It is our hope that the catalog will continue to provide a useful resource for researchers studying galaxy cluster mergers from both the perspective of current observations as well as making predictions for future missions. 

Recently, a similar online catalog of simulated cluster observations been released by \citet{rag16}, which includes 2D maps of relevant quantities, mock X-ray observations, and capabilities for interactive analysis.

\acknowledgements
The authors thank Eric Mandel (CfA) for tips on using JS9 and for critical bug fixes and enhancements that were provided to make it work with the data in the catalog, and Joseph DePasquale (CfA) for generating the home page image. JAZ acknowledges support through Chandra Award Number G04-15088X issued by the Chandra X-ray Center, which is operated by the Smithsonian Astrophysical Observatory for and on behalf of NASA under contract NAS8-03060. KK was funded in part by the Gordon and Betty Moore Foundation's Data-Driven Discovery Initiative through Grant GBMF4561. This work was supported in part by NSF AST-1412768 and by the facilities and staff of the Yale Center for Research Computing.

{}

\end{document}